\documentclass[11pt,fleqn,twoside]{article}
\usepackage{latexsym}
\makeatletter
\newcommand{\prava}{\footnotesize\it
\begin{flushright}
\begin{minipage}{6cm}
Copyright \copyright 1998 by R. Leandre
\end{minipage}
\end{flushright}}

\newcommand{\name}[1]{\begin{flushleft}
                       \LARGE \bf #1
                       \end{flushleft}\vspace{-3mm}}

\newcommand{\Author}[1]{\begin{flushleft}
                       \it #1 \end{flushleft}}

\newcommand{\Adress}[1]{\begin{flushleft}
                       \it #1 \end{flushleft}}

\newcommand{\Date}[1]{\begin{flushleft}
                      \small  \it #1 \end{flushleft}}

\newcommand{\ehkol}{Author \ name}
\newcommand{\ohkol}{Article \ name}
\renewcommand{\@evenhead}{
\hspace*{-3pt}\raisebox{-15pt}[\headheight][0pt]{\vbox{\hbox to \textwidth
{\thepage \hfil \ehkol}\vskip4pt \hrule}}}
\renewcommand{\@oddhead}{
\hspace*{-3pt}\raisebox{-15pt}[\headheight][0pt]{\vbox{\hbox to \textwidth
{\ohkol \hfil \thepage}\vskip4pt\hrule}}}
\renewcommand{\@evenfoot}{}
\renewcommand{\@oddfoot}{}

\newcommand{\be}{\begin{equation}}
\newcommand{\ee}{\end{equation}}
\newcommand{\ba}{\hspace*{-5pt}\begin{array}}
\newcommand{\ea}{\end{array}}

\newcommand{\ds}{\displaystyle}
\makeatother
\begin{document}

\renewcommand{\theequation}{\arabic{section}.\arabic{equation}}
\renewcommand{\thesection}{\arabic{section}}
\setcounter{page}{23}

\thispagestyle{empty}

\renewcommand{\ehkol}{R. Leandre}
\renewcommand{\ohkol}{Stochastic Cohomology}

\begin{flushleft}
\footnotesize \sf
Journal of Nonlinear Mathematical Physics \qquad 1998, V.5, N~1,\ 
\pageref{leandre-fp}--\pageref{leandre-lp}.\hfill {\sc Article}
\end{flushleft}

\vspace{-5mm}

{\renewcommand{\footnoterule}{}
{\renewcommand{\thefootnote}{}  \footnote{\prava}}

\name{Stochastic Cohomology of the Frame Bundle \\
of the Loop Space}\label{leandre-fp}

\Author{R. L\'EANDRE}

\Adress{Department de Math\'ematiques, Institut Elie Cartan, Universit\'e
de Nancy I,\\
54000 Vandoeuvre-les-Nancy, France\\[1mm]
and\\[1mm]
Department of Mathematics, Max Planck Institut f\"ur Mathematics,\\
D5300, Bonn, Germany\\[1mm]
E-mail: Remi.Leandre@antares.iecn.u-nancy.fr}

\Date{Received January 10, 1997; Accepted September 1, 1997}

\begin{abstract}
\noindent
We study the dif\/ferential forms over the frame bundle of
the based loop space. They are stochastics in the sense that we put over
this frame bundle a probability measure. In order to understand the
curvatures phenomena which appear when we look at the Lie bracket of
two horizontal vector f\/ields, we impose some regularity assumptions
over the kernels of the dif\/ferential forms. This allows us to def\/ine an
exterior stochastic dif\/ferential derivative over these forms.
\end{abstract}

\section*{Introduction}

Let $L_x(M)$ be the based loop space of smooth applications
$\gamma_s$ from the circle into $M$ such that $\gamma_0 = \gamma_1 =
x$. Let $Q \rightarrow M$ be a principal bundle over $M$ with structure
group $G$. $L_e(Q)$ is the set of based loop in $Q$ over the based loop
space of $M$. It is a based loop group bundle whose the structure group is
$L_e(G)$, the based loop group of $G$. If $Q \rightarrow M$ is the frame
bundle, $L_e(Q)$ is the frame bundle of $L_x(M)$: the structure of
$L_e(Q)$ is of the main importance to study string structures (or spin
structures) over the loop space ([8], [9], [38]), and has a deep
place in the understanding the Dirac operator over the loop space
([38]).

Let us suppose that the loop space is simply connected, in order to avoid
all torsion phenomenon. If the loop space is the space of smooth loop,
there is an equivalence between the cohomology with values in $Z$ and
$S^1$ bundles over the loop space. Let us now endow the loop space with
the Brownian bridge measure, if the manifold is supposed riemannian. The
equivalence is not at all clear in the stochastic context: let us clarify
what it means. In the stochastic context, the loop are only continuous.  A
stochastic cohomology of $L_x(M)$ is def\/ined in [27], [29] and [30]
 with values in $C$ or $R$: since $L_x(M)$ is
supposed simply
connected, we can neglect all torsion phenomenon in order to construct a
$S^1$ bundle from a $Z$ closed 2 form over the loop space of f\/inite energy
loops. But we have to choose distinguished  paths in $L_x(M)$
in order to shrink a loop in a constant loop: let $l_t(\gamma)_s$ such a
distinguished path. The law of $l_t(\gamma)_.$ is not absolutely
continuous with respect of the law of $\gamma$. So we have to consider
special type of forms in order to overcome the problem: this avoids to use
a $Z$ stochastic cohomology of the brownian bridge, by considering only
examples.

The goal of this paper is to do a stochastic cohomology of the frame
bundle of $L_x(M)$, to construct the stochastic forms which allow to
consider a string structure over $L_x(M)$. Namely, we have already
constructed stochastic bundles over $L_e(Q)$ by starting from a given
deterministic form over this set, and the goal of this paper is to give a
stochastic meaning to this form [34].

As in [34], we def\/ine a measure over $L_e(Q)$, by putching
together measures in the f\/iber: the f\/iber is a continuous loop group. We
start with the equation in the f\/iber
\be\label{0.1}
 dg_s = dB_s g_s.
\ee
In [34], we have studied the equation
\be \label{0.2}
dg_s = B_s ds g_s .
\ee
We choose this equation in order to ref\/lect the f\/iber structure of $L_e(Q)$,
the only obstacle to the trivialization being the holonomy over a loop in
the basical manifold. Namely, we can consider the Albeverio-Hoegh-Krohn
quasi invariance formulas under the right translation $g_. \rightarrow
g_. K_.$. If $K_.$ is deterministic in $C^1$, the quasi-invariance density
belongs to all the $L^p$ in the f\/irst case, while it belongs only to
$L^1$ in the second case, if $K_.$ is $C^2$.

This allows us to def\/ine a tangent space of $L_e(Q)$ by using an inf\/inite
dimensional connection and to get horizontal vector f\/ields and vertical
vector f\/ields. We meet the following paradoxe: the big dif\/ference between
the Sobolev Calculus over the loop group and the Sobolev Calculus over the
loop space of a riemannian manifold is the following: in the f\/irst case,
the tangent vector f\/ields are stable by Lie Bracket, in the second case no.
Apparently, if we follow this remark, we have to separate the treatment
of the horizontal component and of the vertical component of a form, in
order to def\/ine a stochastic exterior derivative over $L_e(Q)$. Let us
recall namely that, in order to def\/ine some cohomology groups over the
loop spaces, we have imposed in [27] some regularity assumptions
over the kernels of the associated forms, in order to simplify the
treatment of the anticipative Stratonovitch integrals which appears in
the def\/inition of the exterior stochastic dif\/ferential. These
conditions lead to needless complications in the case of loop groups
[15].  But in our situation, we cannot neglect the curvature phenomena which
appear: we are obliged to treat the horizontal and the vertical components
in the same manner, in order to def\/ine some stochastic cohomology groups
of $L_e(Q)$. The Carey-Murray [38] form is closed for this stochastic
cohomology (If the f\/irst Pontryaguin class of $Q$ vanishes), because it is
a mixture between a basical iterated integral and the canonical 2 form
over a loop group: this gives the second aspect of the construction of the
string bundle in our stochastic situation.

Moreover, this Calculus depends apparently of the connection over the
frame bundle $L_e(Q) \rightarrow L_x(M)$. But we show that the
functional spaces which are got with some regularity assumptions over
the kernels are independant of this connection.

}

\setcounter{section}{1}
\setcounter{equation}{0}

\section*{Stochastic cohomology of the loop space of the bundle}

Let $Q \rightarrow M$ be a principal bundle with a compact connected
structural Lie goup $G$.  We suppose that $M$ is endowed with a
Riemannian metric: there exists a heat semi-group over $M$ and a
brownian bridge measure $dP_{1,x}$ associated to the riemannian
metric. It is a measure over the based continuous loop space.\par
Over $G$, we consider the following stochastic dif\/ferential equation:
\be \label{1.1}
dg_s = dB_s g_s;\qquad g_0 = e,
\ee
where $B_s$ is a brownian motion independant of the law of the loop
$\gamma$ over $M$ over Lie $G$.

We get a law $Q$ which can be desintegrated over the pinned path space
of paths in the group joining $e$ to $g$ ([17], [2], [3]).
We get a space of continuous paths in $G$ $L_g(G)$ endowed with a law
$Q_g$. The non pinned based path group is denoted $P(G)$.

We put over the bundle $Q \rightarrow M$ a connection $\nabla^Q$:
$\tau_s^Q$ the parrallel transport for a loop $\gamma_s$ is therefore
almost surley def\/ined for the connection $\nabla^Q$. We denote by
$L_e(Q)$ the space of loop $q_.$ in $Q$ such that $q_s = \tau_s^Q g_s$,
$g_1 = (\tau_1^Q)^{-1}$. We get the following commutative diagramm [38]:
\be\label{1.2}
 \matrix{L_e(Q)& \rightarrow &P(G) \cr \downarrow&& \downarrow
\cr
L_x(M) & \rightarrow &G \cr}.
\ee
The map from P(G) to $G$ is the map which to $g_.$ associates $g_1$.
The map from $L_e(Q)$ to $L_x(M)$ is the projection map. The map $f$
from
$L_x(M)$ to $G$ is the map which to a stochastic loop $\gamma_.$
associates $(\tau_1^Q)^{-1}$. The map from $L_e(Q)$ to $P(G)$ is the map
which to $q_.$ associates $g_.$. It is nothing else than $f^*$.\par
Over $L_e(Q)$, we put the measure:
\be \label{1.3}
dP_{tot} = dP_{1,x} \otimes dQ_{(\tau_1)^{-1}}.
\ee

Let us analyze a little bit more the $L_e(G)$ bundle $P(G) \rightarrow G$.
If $g_1 \in G_i$ is a small open neighborhood of $G$, we can choose a
section $g_{i,s}(g_1)$ of this bundle which is jointly smooth in $s$ and
in $g_1$. It checks the following property: $g_{i,0}(g_1)=e$;
$g_{i,1}(g_1) = g_1$; $g_{i,s}(g_1) \in G$. This means that the transition
functions of $P(G)$ can be choosen to take their values in the smooth
based loop space og $G$, $L_e^{\infty}(G)$. Since $G$ is a compact
manifold, we can choose a connection over the bundle $P(G) \rightarrow
G$ whose the structural group is reduced to $L_e^{\infty}(G)$. Let us call
$\nabla^{\infty}$ this connection: if $g_1 \in G_i$, the connection one
form is a smooth path in the Lie algebra of $G$ starting from 0 and
arriving at 0 $K_{i,s}(g_1)$, which depends smoothly from $g_1 \in G_i$
and which is a one form in $g_1$.

The obstruction to trivialize $L_e(Q)$ over $L_x(M)$ lies in
$(\tau_1^Q)^{-1}$: if $(\tau_1^Q)^{-1} \in G_i$, there is a local slice of
$L_e(Q)$ which is $g_{i,.}((\tau_1^Q)^{-1})$. We look at the left
transformation $g_. \rightarrow (g_{i,.}((\tau_1^Q)^{-1}))^{-1}g_.$. Modulo
this transformation, the bridge between $e$ and $(\tau_1^Q)^{-1}$ is
transformed into the bridge between $e$ and $e$. Let us recall namely the
purpose of the quasi-invariance formula from Albeverio-Hoegh-Krohn
[4]: if $k_s$ is a deterministic $C^1$ path in the group $G$, the law
of $g_.k_.$ and the law of $k_.g_.$ are quasi-invariant with respect to the
law of (1.1). Moreover the density of quasi-invariance belong to all the
$L^p$ and can be desintegrated along the appropriate bridge. We denote by
$J_r(k)$ and by $J_l(k)$ the right quasi-invariance density and the left
quasi-invariance density [4], [17].

Therefore if $(\tau_1^Q)^{-1} \in G_i$:
\be \label{1.4}
dP_{tot} = dP_{1,x} \otimes J_l(g_{i,}((\tau_1^Q)^{-1})) dQ_e.
\ee
$J_l(g_{i,.}((\tau_1^Q)^{-1})$ belongs to all the $L^p$ and is bounded in
$L^p$ when $(\tau_1^Q)^{-1}$ describes $G_i$. (1.4) produces a stochastic
trivialization of our bundle.

Let us recall that a vector f\/ield over $L_x(M)$ is given by [7], [20]
\be \label{1.5}
X_t = \tau_t H_t\qquad  X_0 = X_1 = 0,
\ee
where $\tau_t$ is the parallel transport associated to the Levi-Civita
connection and $H_.$ is a f\/inite energy path in $T_x$. We choose as
Hilbertian norm of $X_.$ the norm
\be \label{1.6}
\| X \|^2 = \int\limits_0^1 \| H'_s \|^2 ds = \| \,H\|^2.
\ee

Let us recall that a right vector f\/ield over $L_e(G)$ is given by $X_t^r =
g_t K_t$ where $K_t$ is a f\/inite energy path with end points equal to 0 in
the Lie algebra of $G$ which checks $\ds \int\limits_0^1 \|
K'_s\|^2ds =  \| K\|^2<
\infty$. A left vector f\/ield over $L_e(G)$ is given by $K_s g_s = X_s^l$
where $K_s$ checks the same condition (See [34]).

We pullback the connection $\nabla^{\infty}$ to be a connection over the
stochastic bundle $L_e(Q) \rightarrow L_x(M)$. If $s \rightarrow
K_{i,s}(g_1)(dg_1)$ is the connection form for $g_1 \in G_i$, the
connection form of the pullback connection still denoted $\nabla^{\infty}$
is $s \rightarrow K_{i,s}((\tau_1^Q)^{-1})(\langle
d(\tau_1^Q)^{-1},.\rangle)$. For that,
we recall that:
\be\label{1.7}
\langle d\tau_1^Q,X\rangle = \tau_1^Q \int\limits_0^1
(\tau_s^Q)^{-1}R^Q(d\gamma_s,X_s)\tau_s^Q,
\ee
where $R^Q$ is the curvature tensor of $\nabla^Q$.

We def\/ine as tangent space of the total space $L_e(Q)$ the orthonormal
sum of the horizontal vector f\/ields and vertical vector f\/ields. In the
trivialization given by (1.4), the horizontal vector f\/ields are given by:
\be \label{1.8}
X^H(H)_s = \tau_s H_s - K_{i,s}((\tau_1^Q)^{-1})\langle
d(\tau_1^Q)^{-1},X \rangle
g_{i,s}
\ee
and the vertical vector f\/ields are given by $q_s K_s = X^V(K)_s$. We
choose as Hilbert norm of $X^H(H)$ the quantity $\|H\|^2$ and of $X^V(K)$
the quantity $\|K\|^2$. These vector f\/ields are consistently def\/ined (See
[34]).
Let us recall the following theorem [34]:

\medskip

\noindent
{\bf Theorem I.1.} {\it Let us consider a cylindrical functional
$F(q_{s_1},\ldots,q_{s_r})$ over $L_e(Q)$. Then there exists a
functional ${\rm div}\, X^H(H)$ and a functional
${\rm div}\, X^V(K)$ which belong to all the $L^p$ such
that for deterministic $H$ and $K$:
\be \label{1.9}
E_{tot}\left[\langle dF, X^H(H)\rangle\right] = E_{tot}\left[F
 \,{\rm div}\,X^H(H)\right]
\ee
and such that
\be \label{1.10}
E_{tot}\left[\langle dF, X^V(K)\rangle \right] = E_{tot}\left[F \,{\rm
div}\, X^V(K)\right].
\ee
}
\medskip

Let us introduce over $T^H(L_e(Q))$ and $T^V(L_e(Q))$ a connection:
\be \label{1.11}
\nabla X^V(K) = X^V(\nabla K)
\ee
and
\be \label{1.12}
\nabla X^H(H) = X^H(\nabla H),
\ee
where $\nabla K_s$ is the H-derivative of $K_s$ in the f\/ixed Lie algebra
of $G$ and $\nabla H_s$ is the H-derivative in the f\/ixed tangent space at
$x$ of $M$. The integration by parts (1.9) and (1.10) allow to def\/ine
consistently these derivatives.

If $K_s = \sum k^i_s k_i$ where $k_i$ is an orthonormal basis of the Lie
algebra of $G$, we get:
\be \label{1.13}
\nabla_XK_s = \sum <dk^i_s, X> k_i.
\ee
The same holds for $H_s = \sum h^i_s e_i$ where $e_i$ is a f\/ixed basis of
the tangent space of $M$ at $x$.

Let us consider a n cotensor $\omega$ over $L_e(Q)$. Let us recall that
$\nabla \omega$ is def\/ined as follows:
\be \label{1.14}
\hspace*{-9.3pt}\nabla \omega (X_1,..,X_{n+1}) = \langle d(\omega (X_1,\ldots,
X_n)), X_{n+1}\rangle  - \sum_{i=1}^n \omega (X_1,\ldots,
\nabla _{X_{n+1}}X_i,\ldots,X_n).
\ee
This allows us to def\/ine iteratively the $k$ covariant derivative of a $n$
form $\sigma$. Let us describe a bit the situation: a $n$ form is a $n$
antisymmetric tensor over the tangent Hilbert space of $q_.$, which has a
priori two types of behaviour:

\begin{enumerate}
\item[--] The horizontal contribution.

\item[--] The vertical contribution.

\end{enumerate}

These contributions have two dif\/ferent behaviours:
\be\label{1.15} \hspace*{-19.7pt}
\ba{l}
\ds \sigma \left(X^H(H_1),\ldots,X^H(H_n),
X^V(K_1),\ldots,X^V(K_m)\right)  =\\[3mm]
\ds  \int\int\sigma^{n,m}(s_1,\ldots,s_n;t_1,\ldots,t_m)
H'_{1,s_1}\ldots H'_{n,s_n}K'_{1,t_1}\ldots K'_{m,t_m}ds_1
\ldots ds_n dt_1\ldots dt_m,
\ea
\ee
where $\sigma^{n,m}$ is a kernel which checks the antisymmetric
conditions due to the antisymmetric conditions over $\sigma$. The
covariant derivatives of $\sigma$ have too two dif\/ferent contributions
which are due to the vertical and horizontal vector f\/ields. In order to
simplify the exposure, we won't do in the formulas the dif\/ference
between the two type of contributions: a form $\sigma$ is given by
kernels $\sigma (s_1,\ldots,s_n)$ whose the covariant derivatives with
respect to the connection $\nabla$ are given by kernels $\sigma
(s_1,\ldots,s_n;t_1,\ldots,t_k)$. Moreover $\int
\sigma(s_1,\ldots,s_n; t_1,\ldots,t_k) ds_i= 0$
and $\int \sigma(s_1,\ldots,s_n; t_1,\ldots,t_k) dt_j = 0$ since we work over
the loop space.

Let us def\/ine the Nualart-Pardoux constants of $\sigma$. Let  $K$ be a
connected component of $[0,1]^n \times [0,1]^k$ where we had removed the
diagonals. We def\/ine the f\/irst Nualart-Pardoux constant as
$C(p,n,k)(Q)$ by the smallest constant such that:
\be \label{1.16}
\ba{l}
\ds \|\sigma (s_1,\ldots,s_n;t_1,\ldots,t_k)-\sigma (s'_1,\ldots,
s'_n;t'_1,\ldots,t'_k) \|_{L^p}
\leq \\[3mm]
\ds \qquad \qquad \qquad C(p,n,k)(Q)
\left(\sum \sqrt{\mid s_i-s'_i\mid }+ \sum \sqrt{\mid t_i -
t'_i \mid}\right)
\ea
\ee
over any $K$.

The second Nualart-Pardoux constan $C'(p,n,k)(Q)$ is the smallest one such
that for all $s_i$ and all $t_j$:
\be \label{1.17}
\| \sigma (s_1,\ldots,s_n;t_1,\ldots,t_k) \|_{L^p} \leq C'(p,n,k)(Q).
\ee

\medskip

\noindent
{\bf Def\/inition I.2.} {\it A $n$ form is said smooth in the
Nualart-Pardoux sense if the collection of $C(p,n,k)(Q)$ and $C'(p,n,k)(Q)$
is f\/inite.}

\medskip

 We have a theorem whose the proof is the analoguous of the
proof of the theorem I.2 of [27].

\medskip

\noindent
{\bf Theorem I.3.} {\it If $\sigma$ is a $n$ form which is smooth in the
Nualart-Pardoux sense and if $\sigma'$ is a $n'$ form which is smooth in
the Nualart-Pardoux sense, $\sigma \wedge \sigma'$ is a $n+ n'$ form
which is still smooth in the Nualart-Pardoux sense.}

\medskip

Over $P(G)$, we can consider the brownian motion measure: $g_1$ is
free. The tangent space of a path $g_s$ is given by the set of vector of the
shape $g_s K_s = X_s$ where $K_0$ is equal to 0 and $K_1$ is free. It is
endowed with the Hilbert structure $\int\limits_0^1 \|K'_s \|^2ds$. We can
repeat the previous considerations and give the def\/inition of a form
which is smooth in the Nualart-Pardoux sense over $P(G)$: its
Nualart-Pardoux constants are called $C(p,n,k)(G)$ and $C'(p,n,k)(G)$. We
choose the same connection than in $Q$ for the def\/inition of iterated
covariant derivatives of a form over $P(G)$.

In the same way, over $L_x(M)$, we can consider the
brownian bridge measure. The tangent space of a loop is the space of
$\tau_s H_s$, $H_0 = H_1 = 0$ and we choose the Hilbert structure
$\int\limits_0^1 \|H'\|^2 ds$. We choose the same connection as
before in order
to iterate the covariant derivatives of a form. If $\sigma$ is a $n$ form,
we can def\/ine its Nualart-Pardoux constant $C(p,n,k)(M)$ and
$C'(p,n,k)(M)$.

\medskip

\noindent
{\bf Theorem I.4.} {\it Let $\sigma(M)$ be a n form over $L_x(M)$ which
belongs to all the Nualart-Pardoux spaces. Then $\pi^* \sigma (M) =
\sigma (Q)$ is a n form over $L_e(Q)$ which belongs to all the
Nualart-Pardoux spaces.}

\medskip

The proof is clear: the Nualart-Pardoux constants are the same. It is not
the same for the next theorem:

\medskip

\noindent
{\bf Theorem I.5.} {\it Let $\sigma(G)$ be a $n$ form over $P(G)$ which
belongs to all the Nualart-Pardoux spaces. Then $(f^*)^* \sigma (G) =
\sigma (Q)$ is a $n$ form which belongs to all the Nualart-Pardoux
spaces over $L_e(Q)$.}

\medskip

{\bf Proof.} Since the functional $\gamma_. \rightarrow h(\tau_1^Q)$
belongs to all the Nualart-Pardoux spaces over $L_x(M)$ if $h$ is
smooth, because the covariant derivatives of $\tau_1^Q$ are given by
iterated integrals (See (1.7)), we can work in a region where $L_e(Q)$ is
trivial, by using a partition of unity over $G$ associated to the $G_i$.
$L_e(Q)$ is locally a product, and we can speak of basical and (left
or right) vertical vector f\/ields. By the lemma A.2. of the appendix, the
Nualart-Pardoux norms in terms of basical and right vertical vector
f\/ields are equivalent to the Nualart-Pardoux norms in term of the right
vertical vector f\/ields and the horizontal vector f\/ields.

It remains to show that $(f^*)^*\sigma$ has locally Nualart-Pardoux
constants for the basical and the right vertical vector f\/ields which are
f\/inite.

The vertical derivative are given by the vertical derivatives over a
pinned path group: the basical one are view by using the derivatives of
$g_{i,.}((\tau_1^Q)^{-1})$ which check the Nualart-Pardoux conditions,
because $(\tau_1^Q)$ checks the Nualart-Pardoux conditions and the
$g_{i,s}$ are smooth in $\tau_1^Q$ and $s$ together. It remains to solve
the problem that we don't consider the form $\sigma$ over $P(G)$  but
the form over $L_{(\tau_1^Q)^{-1}}(G)$ isomorphic to $L_e(G)$ by the
map $g_. \rightarrow g_{i,.}((\tau_1^Q)^{-1})g_.$ This leads to the
vector f\/ield $\langle dg_{i,.}((\tau_1^Q)^{-1},X\rangle g_.$ which is a left vector f\/ield
over $P(G)$. But a map which checks the Nualart-Pardoux conditions
over $P(G)$ for right vector f\/ields checks still the Nualart-Pardoux
conditions for left vector f\/ields (See lemma A.3): if we had consider as
trivialization the couple of $L_x(M) \times P(G)$ with the trivialization
map $(\gamma_., g_.) \rightarrow (\gamma_.,
g_{i,.}((\tau_1^Q)^{-1})g_.)$, the proof would be f\/inished. But a map over
$P(G)$ can be reduced into a map over $L_e(G)$ if it satisf\/ies the
Nualart-Pardoux conditions (See [27] beginning of the chapter II for
the Riemannian case which is more complicated). So if $\sigma
(\gamma_., g_.$) checks the Nualart-Pardoux conditions over $L_x(M)
\times P(G)$, it checks still the Nualart-Pardoux conditions over
$L_x(M) \times L_e(G)$: the map which is associated is the map
$(\gamma_., g_.) \rightarrow ( g_1)$ which gives $L_x(M) \times L_e(G)$
as a f\/inite codimensional manifold of $L_x(M) \times P(G)$.
\hfill $\Box$

\bigskip

\noindent
{\bf Example.} Let $Q \rightarrow M$ a principal bundle with structure
group $G$ which is supposed simply connected simply laced [39]. Over
$P(G)$ we consider the form which at the level of the Lie algebra of
$P(G)$ is equal for right vector f\/ields to:
\be \label{1.18}
c(X,Y) = {1 \over 8 \pi^2} \int\limits_0^1\langle X_s,dY_s\rangle -
\langle Y_s, dX_s\rangle .
\ee
The form $(f^*)^*c$ satisf\/ies over $L_e(Q)$ the Nualart-Pardoux
conditions by the theorem I.5.

Let $\mu$ be the form over $L_x(M)$:
\be \label{1.19}
\mu = {1 \over
8\pi^2}\int\limits_{0<u<s<1}\langle
(\tau_s^Q)^{-1}R^Q(d\gamma_s,.)\tau_s^Q \wedge
(\tau_u^Q)^{-1}R^Q(d\gamma_u,.) \tau_u^Q\rangle .
\ee
It satisf\/ies the Nualart-Pardoux conditions over $L_x(M)$. Therefore
$\pi^* \mu$ satisf\/ies the Nualart-Pardoux conditions over $L_e(Q)$. Let
$\nu$ be a form such that $d\nu = p_1^Q$, the f\/irst Pontryaguin class of
the bundle $Q$ which is supposed to be zero in class. Let $\tau (\nu)$
be the two form over $L_x(M)$
\be \label{1.20}
\tau (\nu) = \int\limits_0^1 \nu (d\gamma_s,.,.).
\ee
It satisf\/ies to the Nualart-Pardoux conditions. Therefore the
Carey-Murray two form [8] $F_Q = (f^*)^*c - \pi^*(\mu + \tau (\nu ))$
satisf\/ies to the Nualart-Pardoux conditions over the big space
$L_e(Q)$.

\medskip

\noindent
{\bf Theorem I.6.} {\it The space of $n$ form which are smooth in the
Nualart-Pardoux sense is independent of the connection
$\nabla^{\infty}$.}
\medskip

{\bf Proof.} Let us consider two connections $\nabla^{\infty}$ and
$\nabla'^{\infty}$. After using a partition of unity associated to the open
neighborhoods $G_i$, since $\tau_1^Q$ satisf\/ies to the
Nualart-Pardoux conditions, we come back to a situation where the bundle
$L_e(Q)$ is trivialized. The Nualart-Pardoux constants of $\sigma$ with
respect of the connection $\nabla^{\infty}$ can be estimated in terms of
the basical Nualart-Pardoux constants, which can be estimated
themselves in terms of the Nualart-Pardoux constants with respect of the
connection $\nabla'^{\infty}$ (See lemma A.2). Therefore the result.
\hfill $\Box$

\medskip

Let us recall that the exterior derivative of a $n-1$ form $\sigma$ is
def\/ined as follows:
\be \label{1.21}
\ba{l}
\ds  d\sigma (X_1,\ldots,X_n) = \sum (-1)^{i-1}\langle d\sigma (X_1,
\ldots,X_{i-1},X_{i+1}\ldots X_n),X_i\rangle \\[3mm]
\ds \qquad  +\sum_{i<j}(-1)^{i+j}\sigma \left([X_i,X_j],X_1,\ldots, X_{i-1},
X_{i+1},\ldots,X_{j-1},X_{j+1},\ldots,X_n\right).
\ea
\ee
Our goal is to def\/ine an exterior derivative over $l_e(Q)$. (1.21) shows
that we need to compute some Lie brackets.

\begin{enumerate}
\item[--]
Let us compute the Lie bracket of two (right) vertical  vector f\/ields. It
is nothing else than $g_t[K^1_t, K^2_t]$ if $X_t^V(K^i) = g_t K^i_t$ for
deterministic process $K^i_t$ in the Lie algebra of $G$.

\item[--]
In order to compute the Lie bracket of two horizontal vector f\/ields, we
work in a local trivialization of $L_e(Q)$. The f\/irst horizontal
vector f\/ild is given by
\be\label{1.22}
X^H(H^1)_t = \tau_t H^1_t - K_{i,t}\left(\langle d(\tau_1^Q)^{-1},
 \tau_.H^1_.\rangle \right) g_t
\ee
and the second one is given by
\be \label{1.23}
X^H(H^2)_t = \tau_t H^2_t - K_{i,t}\left(\langle d(\tau_1^Q)^{-1},
\tau_.H^2_.\rangle \right)g_t
\ee
for deterministic $H^1_t$ and $H^2_t$. We have:
\be \label{1.24}
\ba{l}
\ds [X^H(H^1), X^H(H^2)]_t = \tau_t\int\limits_0^t
\tau_s^{-1}R(d\gamma_s, \tau_s H^2_s) \tau_s H^1_t \\[3mm]
\qquad \ds -\langle d(K_{i,t}(\langle d(\tau_1^Q)^{-1},\tau_.H^1_.
\rangle ), \tau_.H^2\rangle g_t\\[3mm]
\qquad \ds +\langle K_{i,t}(\langle d(\tau_1^Q)^{-1}, \tau_.H^1_.
\rangle )K_{i,t}(\langle d(\tau_1^Q)^{-1},
\tau_.H^2_.\rangle )\rangle g_t\\[3mm]
\qquad  \ds  +\mbox{antisymmetry} =X^H[\tau_.H^1_.,\tau_.H^2_.]
+ R^{\infty}(H^1_.,H^2_.)g_.
\ea
\ee
In other terms, the Lie bracket of horizontal vector f\/ields is not an
horizontal vector f\/ield associated to the generalized vector f\/ield $[\tau_.
H^1_., \tau_. H^2_.]$: some curvature phenomenon appears, which leads to
some extra (left) vertical f\/ields over the f\/iber (and not some (right)
vector f\/ields).

\item[--] The Lie bracket of an horizontal vector f\/ield $X^H(H)$ ($H$
deterministic) and of a vertical vector f\/ield $X^V(K)$ ($K$ deterministic)
is equal to zero.
\end{enumerate}

We are ready to state the following theorem:

\medskip

\noindent
{\bf Theorem I.7.} {\it Let $\sigma$ be an $n$ form which is smooth in the
Nualart-Pardoux sense over $L_e(Q)$. Then $d\sigma$ is a $n+1$ form
which is smooth in the Nualart-Pardoux sense over $L_e(Q)$ and its
Nualart-Pardoux constants can be estimated in terms of the
Nualart-Pardoux constants of $\sigma$.}

\medskip

{\bf Proof.} Only the contribution of the Lie bracket in (1.21) gives any
problem. Since the Lie bracket of two (right) vertical vector f\/ields
is still a (right) vertical f\/ield, only the contribution of the Lie
bracket of two horizontal vector f\/ields put any problem.

We treat f\/irst the contribution of $X^H[\tau_.H^1_., \tau_. H^2_.]$: this
leads to a Stratonovitch integral in $d\gamma_s$: the lemma A.2 of
[27] and more precisely the lemma A.1 of this work allow to show
that this contribution satisf\/ies to the Nualart-Pardoux conditions.

We consider now the contribution of $R_t^{\infty}(H^1_.,H^2_.)g_t$ where
$R_t$ is a process with f\/inite energy in the Lie algebra, which satisf\/ies
to the Nualart-Pardoux conditions because $K_{i,t}$ is smooth in $t$ and
because $\tau_1^Q$ and $\tau_s$ satisf\/y to the Nualart-Pardoux
conditions simultaneously. But
\be \label{1.25}
R_t^{\infty}(H^1_.,H^2_.) g_t =
g_t\left(g_t^{-1}R_t^{\infty}(H^1,H^2) g_t\right)
\ee
which is a generalized (right) vector f\/ield over $L_e(G)$. The proposition
A.4 allows to show that this contribution satisf\/ies still the
Nualart-Pardoux conditions: the only stochastic integral which appears
does not occur from the derivative in time t of $R_t^{\infty}$ but of the
time dif\/ferential element of $g_t^{-1}$ and of $g_t$. \hfill $\Box$

\bigskip

\noindent
{\bf Example.}  Let $\sigma (G)$ be a $n$ form over $P(G)$ which belongs to
all the Nualart-Pardoux spaces over $P(G)$. $d\sigma (G)$ is a $n+1$ form
over $P(G)$ which satisf\/ies to the Nualart-Pardoux conditions. We get
\be \label{1.26}
d(f^*)^*\sigma(G) = (f^*)^*d\sigma(G).
\ee
Namely this property is true if we consider f\/inite energy loop over $Q$,
and ref\/lects some algebraic identities between iterated integrals; these
algebraic identities remain true in the stochastic context.

If we consider a n form $\sigma (M)$ over $L_x(M)$ which satisf\/ies to the
Nualart-Pardoux conditions, $d\sigma (M)$ is a $n+1$ form over $L_x(M)$
which satisf\/ies to the Nualart-Pardoux conditions, and we have clearly:
\be \label{1.27}
\pi^*d\sigma (M) = d\pi^* \sigma(M).
\ee
In particular if the f\/irst Pontryaguin class of the bundle $Q$ is equal to
zero, we can use the result of [8]
\be \label{1.28}
dF_Q = 0
\ee
because $dF_Q$ is equal to zero over the f\/inite energy loop space of $Q$:
$dF_Q$ is given by iterated integrals: these formulas remain true in the
stochastic context.

\renewcommand{\theequation}{a.\arabic{equation}}

\setcounter{equation}{0}

\section*{Appendix: anticipative stratonovitch integrals}

Let us recall the following fact: if $X_s = \tau_s H_s$; $H_0 = H_1 =
0$ is a deterministic vector f\/ield (This means, it corresponds to the
deterministic vector f\/ield $H_s$), we get the following integration by
parts formula, for a cylindrical functional $F$:
\be \label{a.1}
E[\langle dF,X\rangle ] = E[F\, \mbox{div}\, X].
\ee
$\mbox{div}\, X$ is def\/ined by the formula:
\be \label{a.2}
\mbox{div}\, X = \int\limits_0^1\langle \tau_s H'_s,\delta
\gamma_s\rangle  + {1\over 2}
\int\limits_0^1\langle S_{X_s},\delta \gamma_s\rangle .
\ee
$S$ is the Ricci tensor and $\delta $ the Ito integral with respect to the
Levi-Civita connection.

If $K_s g_s$ is a left vector f\/ield ($K_0 = K_1 = 0$; $K_s$
deterministic) over the basical loop group (or any pinned path space
in the group), we get
over the loop group an integration by parts formula analoguous to (a.1),
but this time
\be \label{a.3}
\mbox{div}\,  (K_.g_.) = \int\limits_0^1\langle K'_s, \delta B_s\rangle
\ee
if $g_.$ is given by the equation (1.1).

If $g_s K_s$ is a right vector f\/ield over the basical loop group ($K_.$
deterministic; $K_0 = K_1 = 0$), we get:
\be \label{a.4}
\mbox{div}\, (g_. K_.) =
 \int\limits_0^1\langle g_s K'_s g_s^{-1}, \delta B_s\rangle .
\ee
 We have:

\medskip

\noindent
{\bf Proposition A.1.} {\it Let $u(s,t;u,\tilde{s})$ a random
variable with value
in $T_x(M)$ which satisf\/ies to the Nualart-Pardoux conditions over
$L_e(Q)$, the both type of derivatives included and $s,t,u,\tilde{s}$
included. Then the anticipative Stratonovitch integral:
\be \label{a.5}
\int\limits_s^t\langle \tau_u u(s,t;u \tilde{s}), d\gamma_u\rangle  =
 I(s,t;\tilde{s})
\ee
satisf\/ies to the Nualart-Pardoux conditions, the both type of derivatives
included, and $s, t, \tilde{s}$ included.}

\medskip

{\bf Proof.} We integrate by part in order to compute
$E[(I(s,t;\tilde{s})^p]$
for some even integer $p$. $u(s,t;u,\tilde{s})$ is a vector f\/ield over the
loop space if $\int\limits_s^t u(s,t;u,\tilde{s}) du = 0$. If the
previous inequality
is not checked, we can remove the average of $u$ in order to recognize a
vector f\/ield over the based loop space.

Let us put:
\be \label{a.6}
X_{s,t,\tilde{s}}(u) = \tau_u \int\limits_0^u
u(s,t;v, \tilde{s})dv.
\ee
In the def\/inition of the divergence, we have to add the counterterm ${1
\over 2}\int\limits_0^1\langle S_{X_{s,t,\tilde{s}}(u)},\delta
\gamma_u\rangle $ which is only
apparently an anticipative integral, by integrating by part, in order to
recognize the beginning of a curved Skorohod integral (See [24],
[25]). The second counterterm we have to add is the integral of
the kernels of some H-derivative of $u(s,t;u, \tilde{s})$ in order to
recognize a complete Skorohod integral. For that, we begin by studying
$E[(I(s,t;\tilde{s})^p]$ for the discrete approximation of the anticipative
integral by Riemann sum of \ $\ds {1\over
t_{i+1}-t_1}\int\limits_{t_i}^{t_{i+1}}u(s,t;v,\tilde{s})dv$ \ for a suitable
subdivision of $[0,1]$. We f\/ind a sum of integral over $[s,t]^k$ of polynomial
expression in the derivatives of $u$ and of the derivatives of $\tau_.$ and
of $d\gamma_u$, with possible contraction over the diagonals. Let us
recall that:
\be \label{a.7}
\nabla_X \tau_s = \tau_s \int\limits_0^s \tau_u^{-1}R(d\gamma_u, X_u)
\tau_u.
\ee

We work in a small trivialization of the bundle such that we can speak of
basical derivatives (associated to basical vector f\/ields) instead of
horizontal vector derivatives (associated to horizontal vector f\/ields): the
derivatives which appear after integrating by parts in
$E[(I(s,t;\tilde{s})^p]$ are basical derivatives, and not the horizontal
derivatives which are given by the Nualart-Pardoux norms. This leads
apparently to a problem, which is solved by the Lemma A.2: we postpone
the proof of this lemma later.

\medskip

\noindent
{\bf Lemma A.2.} {\it Let be a local trivialization of the bundle
$L_e(Q))$, such that we can speak of a basical vector f\/ield and of a
(right) vertical vector f\/ield. We can speak of the Nualart-Pardoux
constants of a form $\sigma$
for the basical vector f\/ields and the (right) vertical vector f\/ield.
They can be estimated in terms of the Nualart-Pardoux constants in terms of
horizontal and (right) vertical vector f\/ields. The converse is true.}

\medskip

In the previous discussion, we did not speak of the fact we have removed
to $u(s,t;u,\tilde{s})$ the quantity \ $\ds {1 \over t-s}
\int\limits_s^tu(s,t;u,\tilde{s}) du$ \ as well for the higher derivatives, because some auxiliary terms
which arise from the derivative of the parallel transport can appear. We
f\/ind after this remark a f\/inite sum of integrals over $[s,t]^k$ of
polynomial expressions in the basical derivatives of $u$ with possible
contraction over the diagonals, and some expressions in the basical
derivatives of the parallel transport and of the curvature tensor. It is
possible to divide this integral by a power $k'$ of $t-s$, but we have
always $k'+ p/2 \leq k$: namely the division by $t-s$ appears by an
operation of averaging in order to recognize a vector f\/ield over the loop
space and not from an integration by parts, which leads to at most $p/2$
contractions.

We deduce that the discrete approximation of the integral converges in all
the $L^p$ to $I(s,t;\tilde{s})$, this from the regularity assumption over
the kernels of $u(s,t;u,\tilde{s})$. Moreover $\|I(s,t;\tilde{s})\|_{L^p}^p$
is
a sum of iterated integrals of the basical kernels of $u(s,t;u,\tilde{s}) $
over $[s,t]^k$ with contraction of the basical kernels and half-limits over
the diagonals. Therefore $\|I(s,t;\tilde{s})\|_{L^p}$ is bounded by the
Nualart-Pardoux Sobolev norms of the vector f\/ields by the lemma A.2.

In order to check the regularity assumption in $s,t,\tilde{s}$, we suppose
in order to simplify that $s_1 < \tilde{s}_{1,1} <\ldots <\tilde{s}_{n,1}<t_1$
and that $s_2<\tilde{s}_{1,2}\ldots <\tilde{s}_{n,2}<t_2$.
We splitt the integral
between $s_1$ and $t_1$ into smaller integrals over the intervals
def\/ined by the contiguous time of the subdivision, as it was already done
in the proof of the lemma A.2 of [27].

We consider the integral of $u(s_1,t_1;u,\tilde{s}_1)$ and of
$u(s_2,t_2;u,\tilde{s}_2)$ over $[\tilde{s}_{i,1},\tilde{s}_{i+1,1}]$ $\cap$
$[\tilde{s}_{i,2}, \tilde{s}_{i+1,2}]$. We substract the necessary
counterterm in order to get an anticipative integral over the based loop
space, and we get the expectation of an integral power of it as before. The
main remark is that we remain in the same connected component of the
parameter set outside the diagonals in the integrals which are got. We get
an estimate of the $L^p$ norm in term of $\sqrt{\mid
\tilde{s}_{i,1}-\tilde{s}_{i,2}\mid} + \sqrt{ \mid
\tilde{s}_{i+1,1}-\tilde{s}_{i+1,2}\mid}$. If we integrate outside the
intersections, the distance between the extremities of the considered
intervals is smaller than the sum of the distance between
$\tilde{s}_{i,1}$ and $\tilde{s}_{i,2}$. We get an estimate in terms of the
Nualart-Pardoux constants of the second type (1.17) and $\sqrt{\mid
\tilde{s}_{i,1}-\tilde{s}_{i,2}\mid} + \sqrt{\mid
\tilde{s}_{i+1,1}-\tilde{s}_{i+1,2}\mid}$.

In order to f\/inish the proof, let us precise the ef\/fect of the operation
$\ds u(s,t;u;\tilde{s}) \rightarrow u(s,t;u,\tilde{s}) -
{1 \over t-s} \int\limits_s^tu(s,t;u,\tilde{s})du$ \  over $I$ in order
to get a tangent
vector over the based loop space. It has only the consequence to substract
to the initial anticipative integral the non-anticipative Stratonovitch
integral $\int\limits_s^t<\tau_uC, d\gamma_u>$ for a suitable $C$.

For the derivative of $I$, we deduce from the previous discussion, that we
can take derivative under the sign $\int$, for the vertical and horizontal
vector f\/ields. We conclude by using the fact that:
\be \label{a.8}
\nabla_X d\gamma_s = \tau_s H'_s ds
\ee
if $X_s = \tau_s H_s$.\hfill $\Box$

\medskip

{\bf Proof of the lemma A.2.} The lemma A.2 will be proved if, after
choosing a trivialization which does not af\/fect the Nualart-Pardoux
conditions as we will see later, we prove the following fact: a functional
which belongs to all the Nualart-Pardoux spaces for the vertical
derivatives along right vector  f\/ields $g_.K_.$ belongs to all the
Nualart-Pardoux spaces for the vertical derivatives along (left) vector
f\/ields $K_.g_.$, and its Nualart-Pardoux constants can be estimated in
term of the Nualart-Pardoux constants for the f\/irst type of vector f\/ields.
Namely $K_s$ depends in the def\/inition of an horizontal vector f\/ield only
on $(\tau_1^Q)^{-1}$ and its derivatives, which satisfy to the
Nualart-Pardoux conditions if we consider horizontal vector f\/ields.

A vector f\/ield $g_sK_s$ corresponds to the vector f\/ield $g_sK'_sg_s^{-1}$
over the leading brownian motion. Let us suppose we can get a
prolongation over the path group of our functional which checks the
(right) Nualart-Pardoux conditions and whose the (right) Nualart-Pardoux
 norms over the whole path group can be estimated into the
Nualart-Pardoux norms over the pinned path group. We get therefore a
functional over the leading brownian motion, which belongs to all the
Sobolev spaces. The f\/lat derivative of $g_s$ checks $s$ included the
Nualart-Pardoux conditions: the f\/lat Nualart-Pardoux norms can be
estimated in terms of the Nualart-Pardoux norms over the pinned loop
group. Moreover a (left) vector f\/ield $K_.g_.$ gives the vector
$[K_s,dB_s]+ K'_sds$, which is a generalized f\/lat vector f\/ield. We can
repeat therefore the proof of the theorem A.1 of [27] in order to
conclude. Let us precise this statement.

Let us precise the prolongation: we do in order to simplify as we were
working over the based loop group. We consider a path in $G$ with the
condition that $g_1$ is closed from $e$. We associate to a path $g_s$ the
loop $g_s \exp[-s\,\mbox{Log}\, g_1]$. The vector f\/ield $g_s K_s$
is transformed into the vector f\/ield:
\be \label{a.9}
\ba{l}
\ds g_s \exp[-s \,\mbox{Log}\, g_1]
\left(\exp[s\,\mbox{Log}\, g_1]K_s \exp[-s\,\mbox{Log}\, g_1]\right)
\\[3mm]
\ds \qquad +g_s \exp [-s\, \mbox{Log}_1]\left(\exp [s \,\mbox{Log} g_1]
{\partial \over \partial
g_1}\exp[-s\,\mbox{Log}\,g_1]K_1]\right).
\ea
\ee

Therefore a vector f\/ield $g_s K_s$ is transformed into
\be \label{a.10}
g_s\exp[s\,\mbox{Log}\, g_1] \exp[s\, \mbox{Log}_1]
\left(K_s \exp[-s\,\mbox{Log}\, g_1]
+ {\partial \over \partial g_1}
\exp[-s\,\mbox{Log}\, g_1]K_1\right).
\ee
$g_1$ satisf\/ies to the Nualart-Pardoux conditions and the law of $g_s
\exp[-s\,\mbox{Log}\,C]$ is absolutely continuous with a density
which belongs to
the $L^p$ with respect to the law of $g_s$. It follows than the functional
over the path group:
\be \label{a.11}
F_{tot}(g_.) = F\left(g_.\exp[-.\,\mbox{Log}\, g_1]\right) \phi(g_1).
\ee
where $\phi (g_1)$ is a cutof\/f functional destinated to ensure the
existence of $\mbox{Log}\, g_1$ belongs to all the (right) vertical Nualart-Pardoux
Sobolev spaces, and its vertical Nualart-Pardoux constants can be
estimated by the vertical Nualart-Pardoux constants over the pinned loop
group.

Let us now repeat the scheme of the proof of the theorem A.1. of [27].
Let $\sigma_n = \sigma (B_{u_1},\ldots,B_{u_n})$, $0 = u_1<\ldots<u_n
= 1$, $u_{i+1}-u_i = 1/n$ for a dyadic subdivision of length $2^k$.
Let $F_n =E[F|\sigma_n]$. It is a functional which depends only from
a f\/inite number
of f\/lat variables. Since $dB_s = dg_s g_s^{-1}$, $F_n$  belongs to all the
Sobolev spaces for $g_.$ related to the (left) vector f\/ields $K_.g_.$. The
f\/lat kernel of $F_n$ are given by:
\be \label{a.12}
{1 \over \prod (u_{k_{i+1}}-u_{k_i})} \int \int_{\prod
[u_{k_i},u_{k_{i+1}}]} E[k(t_1,\ldots,t_r)|\, \sigma_n] dt_1\ldots
dt_n.
\ee
$k$ denotes a f\/lat kernel of $F$. $F_n = G_n (g_.)$. We will show that
$G_n$ is a Cauchy sequence for the Sobolev spaces relatively to the left
vector f\/ields $K_.g_.$ (We call them the left Sobolev spaces). The kernel
associated to $G_n$ are Stratonovitch integrals in $dB_s$. We use for this
the formula:
\be \label{a.13}
\Delta B_{u_i} = \int\limits_{u_i}^{u_{i+1}} dg_s g_s^{-1}.
\ee
The derivative of $dg_s$ is  given by $K'_s g_s ds + K_s dg_s$ and the
derivative of $g_s^{-1}$ is given by $-g_s^{-1}K_s$. The kernel of the
derivative of $G_n$ are iterated Stratonovitch integrals with frozen time:
we integrate expressions in the f\/lat derivatives of $F_n$ and algebraic
expressions in $g_s$, $g_s^{-1}$ and $dB_s$ which are non
anticipative.

It remains to pass at the limit: we see that the half limits over the
diagonals of the f\/lat kernels of $F$ appear when we go to the limit.
In order to pass at the limit, there are two procedures as in the proof of
the theorem A.1 of [27]. $\Pi_n$ is the procedure of conditional
expectation over the $\sigma$-algebra $\sigma_n$ and $\chi_n$ the
procedure of averaging. Let $\mbox{Ker}\, F$ be a f\/lat kernel of $F$. The associated
f\/lat kernel of $F_n$ is $\Pi_n \chi_n \,\mbox{Ker}\,F$. We get:
\be \label{a.14}
\Pi_n \chi_n \,\mbox{Ker}\, F - \Pi_{n'} \chi_{n'}\,
 \mbox{Ker}\,F = \Pi_n (\chi_n - \chi_{n'})
\,\mbox{Ker}\,F + (\Pi_n - \Pi_{n'}) \chi_{n'} \,\mbox{Ker}\, F.
\ee
$\Pi _n \,\mbox{Ker}\,F$ satisf\/ies to the Nualart-Pardoux conditions
with the same
constants than $\mbox{Ker}\,F$. We can apply the Kolmogorov lemma.
Let us denote
by $U$ any connected component of the complement of the diagonals. We
get:
\be \label{a.15}
\mbox{Sup}_{U\times U}
{\|E[k(t_1,\ldots,t_r)|F_n]-E[k(t'_1,\ldots,t'_r)|F_n]\|_{L^p} \over
\sum \mid t_i-t'_i\mid^{\alpha}} < \infty
\ee
for a certain $\alpha < 1/2$.

The previous $L^p$ norms are smaller than the $L^p$ norms which are got,
when we don't take any conditional expectation. When we take the $L^p$
norm of the dif\/ference of the Stratonovitch integral of $\Pi_n \chi_n
\,\mbox{Ker}\,F$ and of $\Pi_{n'} \chi_{n'}\, \mbox{Ker}\,F$, we get iterated
integrals without
the stochastic term $dB_s$ with some half limits of the kernels of $F_n -
F_{n'}$ over the diagonals. We splitt it into an expression polynomial in
$(\Pi_n - \Pi_{n'}) \chi_n\,\mbox{Ker}\,F$ and a polynomial expression in
$(\chi_n-\chi_{n'})\Pi_{n'}\,\mbox{Ker}\,F$. The f\/irst type of
expression goes
uniformly to 0 in all the $L^p$ by using the Kolmogorov lemma when $n
\rightarrow \infty$ and $n' \rightarrow \infty$. It is the same for the
second type of expressions, by using the criterium of continuity of the
kernels of $F$. $G_n$ is a Cauchy sequence in the (left) Sobolev
spaces.

Moreover, by the Kolmogorov lemma:
\be \label{a.16}
E\left[k(s_1,\ldots,s_r)|\sigma_n\right] - k(s_1,\ldots,s_r) \rightarrow 0
\ee
uniformly in $(s_1,\ldots,s_r)$ outside the diagonals in all the $L^p$. The
derivatives of $G_n$ tend to the Stratonovitch integral which are got
formally when we replace the f\/lat $dH_s$ by $[K_s,dB_s] + K'_s ds$.

It remains to restricts the functional $G$ over the based loop group as
well as its kernels. It is the purpose of the quasi-sure analysis: we
consider the measure
\be \label{a.17}
f \rightarrow E[G f(g_1)]
\ee
which has a density. The (left) Nualart-Pardoux Sobolev norms for the
pinned loop group are estimated in terms of the (left) Nualart-Pardoux
Sobolev norms over the path group. In order to show that , let us consider
the vector f\/ields $X_.^1 = K_.^1g_.$,\ldots, $X_.^r = K_.^r g_.$. We get the
following integration by parts formula for any integer $p$:
\be \label{a.18}
\ba{l}
E\Bigl[|G(s_1,\ldots,s_r)- \tilde{G}(\tilde{s}_1,\ldots,
\tilde{s}_r)|^p\langle d\langle \ldots \langle df(g_1),X^r
\rangle \ldots\rangle X^1\rangle \Bigr] = \\[2mm]
\hspace*{8cm} E[\xi f(g_1)],
\ea
\ee
where
$\xi$ is a polynomial expression in
$G(s_1,\ldots,s_r)-\tilde{G}(s_1,\ldots, s_r)$ and
its (left) kernels integrated and the divergence of the vector f\/ields $X$
and their derivatives. We apply the lemme A.2 of [27] in order to
conclude.

We have proved the lemma A.2 for functionals: for forms, we associate
to a form over (right) vertical vector f\/ields a form over f\/lat
vector f\/ields, and after we do the transformation $dH_s \rightarrow
[K,s,dB_s] +
K'_sds$ in order to get a form over the (left) vertical vector
f\/ields: we get
Stratonovitch iterated integrals, and we apply the lemma A.2 of [27]
in order to conclude.

The last point it remains to clarify is that the operation of trivialization
in order to come to a product situation has no ef\/fect over the right
vertical Nualart-Pardoux Sobolev norms. If $(\tau_1^Q)^{-1} \in G_i$, we
can f\/ind a mollifer $f(g_1)$ which belongs to all the Nualart-Pardoux
spaces with compact support in a small neighborhood of $G_i$. We put
over the path group
\be \label{a.19}
F_{tot}(g_.) = F\left(g_. \exp[-s\,\mbox{Log}\,[\tau_1^Q g_1]\right) f(g_1).
\ee
We enlarge by this the functional over the vertical pinned path going from
$e$ to $(\tau_1^Q)^{-1}$ to a functional over the total space of the path
group which checks still the Nualart-Pardoux conditions: its right
Nualart-Pardoux norms can be estimated in term of the Nualart-Pardoux
norms of the non extended functional, since $(\tau_1^Q)$ satisf\/ies to the
Nualart-Pardoux conditions. We perform after the transformation:
\be \label{a.20}
F_{tot}(g_.) \rightarrow F_{tot}\left(g_{i,.}((\tau_1^Q)^{-1})g_.\right).
\ee
The left Nualart-Pardoux constants of the new global functional can be
estimated in term of the right Nualart-Pardoux constants of
$F_{tot}(g_.)$. Since the law og $g_i.\left((\tau_1^Q)^{-1}\right)g_.$ is equivalent to
the law of $g_.$ with a density  which belongs to all the $L^p$, we deduce
that the global Nualart-Pardoux constants of
$F_{tot}\left(g_{i.}\left((\tau_1^Q)^{-1}g_.\right)\right)$ can be
estimated in terms of the right
Nualart-Pardoux constants of $F(g_.)$. \hfill $\Box$

\medskip

We had shown too the following lemma:

\medskip

\noindent
{\bf Lemma A.3.} {\it A functional over a trivialization which checks the
(right) Nualart-Pardoux conditions for (right) vertical vector f\/ields over
$L_e(G)$ (or$ P(G)$) checks still the (left) Nualart-Pardoux conditions
for (left) vertical vector f\/ields.}

\medskip

We get the proposition:

\medskip

\noindent
{\bf Proposition A.4.} {\it Let $u(s,t;u\tilde{s})$ a random
application with
values in $Lie G$ which belongs to the (right) Nualart-Pardoux spaces
over the total space, $s, t, \tilde{s}$ included. Let $I(s,t,\tilde{s})$ the
anticipative Stratonovitch integral:
\be \label{a.21}
I(s,t;\tilde{s}) = \int\limits_s^t\langle g_u u(s,t;u,\tilde{s}),
dg_u\rangle .
\ee
$I(s,t;\tilde{s})$ checks the Nualart-Pardoux conditions over the total
space $L_e(Q)$.}

\medskip

{\bf Proof.} We begin to write $dg_u = dB_u g_u$, such that we come back
to a Stratonovitch integral
\be \label{a.22}
\int\limits_s^t\langle \tilde{u}(s,t;u,\tilde{s}), dB_u\rangle ,
\ee
where $\tilde{u}$ checks still the Nualart-Pardoux conditions. We extend
$\tilde{u}(s,t;u,\tilde{s})$ over the path group, such that it checks still
the Nualart-Pardoux conditions over the path group. We use the isometry
given in the beginning of the proof of the lemma A.2. We get $\tilde{u}$
which depends on $B$ and $\gamma$ which checks the Nualart-Pardoux
conditions in $B$ and $\gamma$. We come back to the f\/lat case and to a
f\/lat Stratonovitch integral. We can use the results of [27]:
$I(s,t;\tilde{s})$ extended satisf\/ies to the Nualart-Pardoux conditions in
$B$ and $\gamma$, for the basical vector f\/ield in $\gamma$. this from
the lemma A.2 of [27]. Then $I(s,t;\tilde{s})$ extended satisf\/ies to
the Nualart-Pardoux conditions in $g_.$ and $\gamma_.$, for basical
vector f\/ields in $\gamma$. By the lemma A.2, it satisf\/ies to the
Nualart-Pardoux conditions in $g_.$ and $\gamma_.$ for horizontal vector
f\/ields, which are intrisically def\/ined. \hfill $\Box$

\section*{Acknowledgments}

We thank the warm hospitality of the Max Planck Institut for Mathematics
in Bonn where this work was done.

\label{leandre-lp}


\begin{thebibliography}{99}
\footnotesize

\bibitem{1} Aida S. and  Elworthy D.,
 Dif\/ferential Calculus on path and loop spaces, Preprint.

\bibitem{2} Airault H. and Malliavin P., Quasi sure analysis,
Publication Paris VI, 1991.

\bibitem{3} Airault H. and Malliavin P.,
 Integration on loop groups,  Publication Paris VI, 1991.

\bibitem{4} Albeverio S. and Hoegh-Krohn R.,
The energy representation of Sobolev Lie groups, {\it Compositio
Math.}, 1978, V.36, 37--52.

\bibitem{5}  Albeverio S.,  Ma Z.M. and Rockner M.,
Partition of unity in Sobolev spaces over inf\/inite dimensional state spaces,
{\it J.F.A.}, 1997, V.143, 247--267.

\bibitem{6} Arai A., A general class of inf\/inite dimensional
operators and path representation of their index, {\it J.F.A.}, 1992,
V.105, 342--408.
\bibitem{7} Bismut J.M., Large deviations and the Malliavin Calculus,
{\it Progress in Math.}, V.45,  Birkhauser, 1984.

\bibitem{8} Carey A.L. and Murray M.K., String structure and the path
f\/ibration of a group, {\it  C.M.P.}, 1991, V.141, 441--452.

\bibitem{9} Coquereaux R. and Pilch K., String structure on loop
bundles, {\it  C.M.P.}, 1989, V.120, 353--378.

\bibitem{10} Cruzeiro A. and Malliavin P., Renormalized dif\/ferential
geometry on path space: Structural equation, curvature, {\it J.F.A.},
1996, V.139, 119--181.

\bibitem{11} Driver B., A Cameron-Martin type
quasi-invariance for Brownian motion on compact manifolds, {\it
J.F.A.}, 1992, V.110, 272--376.

\bibitem{12} Elworthy D.,
 Stochastic dif\/ferential equations on manifold,
 L.M.S. Lectures Notes Serie 20,  Cambridge University Press, 1982.

\bibitem{13} Elworthy K.D. and Ma M.Z., Vector
f\/ields on mapping spaces and related Dirichlet forms and dif\/fusions,
Preprint.

\bibitem{14} Enchev O. and Stroock D.W., Towards a riemannian
geometry on the path space over a riemannian manifold, {\it J.F.A.},
1996, V.134, 392--416.

\bibitem{15} Fang S. and Franchi J., De Rham-Kodaira operator on loop
group. Preprint.

\bibitem{16} Gross L., Potential theory on
Hilbert spaces, {\it  J.F.A.}, 1967, V.1, 123--181.

\bibitem{17} Gross L., Logarithmic Sobolev inequalities on a loop
group, {\it  J.F.A.}, 1991, V.102, 268--312.

\bibitem{18} Hsu E.P., Quasi-invariance of the Wiener measure on the
path space over a compact Riemannian manifold, {\it J.F.A.}, 1995,
V.134,  417--450.

\bibitem{19} Ikeda N. and Watanabe S., Stochastic dif\/ferential equations
and dif\/fusion processes, North-Holland, 1981.

\bibitem{20} Jones J. and L\'eandre R.,
$L^p$ Chen forms over loop spaces, In: Stochastic Analysis, Barlow M.,
Bingham N. edi.,  Cambridge University Press. 1991, 104--162.

\bibitem{21}  Jones J.D.S. and  L\'eandre R.,
 A stochastic approach to the
Dirac operator over the free loop space,  To be published in: "Loop spaces".
Sergeev A. edit.

\bibitem{22}  Kusuoka S., De Rham cohomology of Wiener-Riemannian
manifolds, In: Proceedings I.C.M., Kyoto, 1990, Springer.,
1075--1082.

\bibitem{23} L\'eandre R., Strange behaviour of the
heat kernel on the diagonal, In: Stochastic processes, physic and geometry,
S. Albeverio edit., World Scientif\/ic, 1990, 516--528.


\bibitem{24} L\'eandre R., Integration by parts formulas and rotationally
invariant Sobolev Calculus on the free loop space, XXVII Winter School of
theoretical physic, Gielerak R.,  Borowiec A. edit.,  {\it J. of Geometry and
Physics}, 1993, V.11,  517--528.

\bibitem{25} L\'eandre R., Invariant Sobolev Calculus on the free
loop space, {\it Acta Applicandae Mathematicae}, 1997, V.46,
267--350.

\bibitem{26} L\'eandre R.,
Brownian motion over a Kahler manifold and elliptic genera of level N, In:
Stochastic Analysis and Applications in Physics, S\'en\'eor R.,
Streit L. edi.,  Nato ASI serie, 1994, V.449, 193--217.

\bibitem{27} L\'eandre R.,  Cohomologie de Bismut-Nualart-Pardoux et
cohomologie de Hochschild en\-tie\-re, S\'e\-mi\-nai\-re de
Probabilit\'es XXX in honour of P.A. Meyer et J. Neveu L.N.M.,
1626, Az\'ema J., Emery M., Yor M. eds, 1996, 68--100.

\bibitem{28} L\'eandre R., Stochastic Wess-Zumino-Witten model over a
symplectic manifold, {\it   Journal of Geometry and
Physics}, 1997,  V.21,  307--336.

\bibitem{29} L\'eandre R., Brownian cohomology of an homogeneous manifold,
Proceedings of the Taniguchi conference,  K.D~ Elworthy,  S.~Kusuoka,
 I.~Shigekawa edit., World Scientif\/ic, 1997, 305--348.

\bibitem{30} L\'eandre R., Stochastic Moore loop space, In: Chaos:
the interplay between stochastic and deterministic behaviour,
 Garbaczweski P. edit., {\it  Lecture Notes in Physics}, 1995, V.457,
 479--502.

\bibitem{31} L\'eandre R., String structure over the brownian bridge,
Preprint.

\bibitem{32}
L\'eandre R., Hilbert space of spinors f\/ields over the free loop
space, {\it Reviews in Mathematical Physics}, 1997, V.9.2, 243--277.

\bibitem{33}  L\'eandre R., Cover of the brownian bridge and
stochastic symplectic action,  To be published in {\it Reviews in Mathematical
Physics}.

\bibitem{34} L\'eandre R.,  Stochastic gauge transform of the
string bundle,  To be published in {\it Journal of Geometry and Physics}.

\bibitem{35} L\'eandre R., Stochastic Wess-Zumino-Witten model for the
measure of Kontsevitch, Preprint.

\bibitem{36} L\'eandre R. and Norris J.,
Integration by parts and Cameron-Martin formulas for the free path space
of a compact Riemannian manifold,  S\'eminaire de
Probabilit\'es XXXI,  L.N.M.,  1655, 1997, 16--24.

\bibitem{37} L\'eandre R. and Roan S.S., A stochastic approach to the
Euler-Poincar\'e number of the loop space of a developable orbifold,
{\it  J. Geometry and Physics}, 1995, V.16, 71--98.

\bibitem{38} Mac Laughlin D., Orientation and string structures on
loop spaces, {\it Pac. J. Math.}, 1992, V.155,  143--156.

\bibitem{39} Pressley A. and Segal G., Loop groups, Oxford University
Press, 1986.

\bibitem{40} Shigekawa I., Transformations of Brownian motion on a
Riemannian symmetric space, {\it Z.W.}, 1984, V.65,  493--522.

\bibitem{41} Shigekawa I., Dif\/ferential Calculus on a based loop
group, Preprint.

\bibitem{42} Taubes C., $S^1$ action and elliptic genera, {\it
C.M.P.},  1989, V.122, 455--526.

\bibitem{43} Witten Ed., The index of the Dirac operator in loop
space, In: Elliptic curvex and modular forms in algebraic topology,
Landweber edit.,  L.N.M.  1326.  Springer, 1988, 161--181.


\end{thebibliography}
\end{document}